\documentclass[a4paper,12pt]{article}
\pdfoutput=1

\usepackage{jheppub} 

\usepackage[T1]{fontenc} 
\usepackage{graphicx}
\usepackage[latin1]{inputenc}

\title{\boldmath Pre-Equilibrium Radial Flow from Central Shock-Wave Collisions in $AdS_5$}

\author[a]{Paul Romatschke}
\author[b]{and J. Drew Hogg}

\affiliation[a]{Department of Physics, 390 UCB, University of Colorado, Boulder, CO 80309-0390, USA }
\affiliation[b]{Department of Astronomy, University of Maryland, College Park, MD 20742, USA}

\emailAdd{paul.romatschke@colorado.edu}

\abstract{Using gauge/gravity duality, central ultrarelativistic nucleus-nucleus collisions are modelled as collisions of shock waves in five-dimensional asymptotic $AdS$ space. For early times after the collision, it is possible to analytically match the metric from the past to the future light-cone. This allows extraction of the pre-equilibrium energy-momentum tensor of the strongly coupled, large $N$ gauge theory. For central collisions, this allows qualitative statements concerning the build-up of radial flow at mid-rapidity 
in $AA$ and $pA$ collisions.
We find that the early-time radial flow buildup is identical to that expected from ideal hydrodynamics with an entropy density proportional to the square root of the product of the matter densities in the individual "nuclei".}

\begin{document}
\maketitle

\section{Introduction}

A long-standing problem in the field of high energy nuclear collisions has been trying to understand the precise mechanism and timing of equilibration after the collision of the incoming nuclei. The success of the heavy-ion program at the Relativistic Heavy Ion Collider (RHIC) \cite{Adcox:2004mh,Back:2004je,Arsene:2004fa,Adams:2005dq} and the Large Hadron Collider (LHC) \cite{Aamodt:2010pa,Aad:2010bu,Chatrchyan:2011sx} has further provided motivation to study this difficult regime, where one needs to describe the real time evolution of non-perturbative coupled quantum field theory. Previous work on the pre-equilibrium regime in nuclear collisions can broadly be classified in two branches: weak-coupling (perturbative) techniques to solve Quantum-Chromodynamics (QCD) \cite{Baier:2000sb,Bjoraker:2000cf,Arnold:2004ti,Xu:2004mz,Rebhan:2004ur,Romatschke:2005pm,Berges:2008zt,Kurkela:2011ub,Dusling:2012ig} and gauge/gravity duality to solve strongly coupled gauge theories different from QCD\cite{Janik:2005zt,Lin:2006rf,Kovchegov:2007pq,Gubser:2008pc,Albacete:2008vs,Grumiller:2008va,AlvarezGaume:2008fx,Chesler:2008hg,Chesler:2010bi,Wu:2011yd,Balasubramanian:2011ur,Mateos:2011tv,Kiritsis:2011yn} . In the former setup, particle-like degrees of freedom are weakly coupled to the classical field background (such as that from the Color-Glass-Condensate) and are known to give rise to a kind of plasma instability \cite{Mrowczynski:1993qm}. At the time of writing, the actual equilibration (transition to hydrodynamic behavior) has not been observed in realistic (e.g. longitudinally expanding) simulations, but lots of progress has been made in toy model systems. 

This work will follow the second, strong-coupling approach, where one permits oneself to trade the non-abelian gauge theory of physical interest (QCD) with a different (non-abelian) gauge theory ${\cal N}=4$ SYM. In this case, using the dictionary from gauge/gravity duality \cite{Son:2007vk}, one can map the energy-distribution of a fast moving nucleus to a gravitational shock wave in asymptotically $AdS$ space in five-dimensions (cf.~\cite{Gubser:2008pc}). Using machinery from relativity, one can draw on previous work that has established the form of the line element for given (energy-) density profiles of nuclei, a lower bound on the total entropy produced in the collision of two of the shock waves. Moreover, using techniques from numerical relativity on the collision of black holes in various dimensions, the actual equilibration of the gauge theory from a far-from equilibrium state to hydrodynamics has been observed in numerical simulations with a high degree of symmetry (the nuclei where assumed to be translationally invariant in the plane transverse to the collision axis) \cite{Chesler:2010bi,Wu:2011yd}. However, these simulations could not provide information on the pre-equilibrium dynamics in the transverse plane, which is of interest because it could potentially lead to observable effects in nuclear collision experiments at RHIC and the LHC. The aim of the present article is to provide the foundation to lift this shortcoming by providing the metric shortly after the head-on collision of two gravitational shock waves, including the extraction of the early-time gauge-theory energy-momentum tensor. By the nature of the early-time series employed, the results obtained will be quantitatively reliable only at mid-rapidity and close to the boundary of AdS space. Nevertheless, we presume that using these results in conjunction with recent numerical advances in solving the Einstein equations in $AdS_5$ (cf.~\cite{vanderSchee:2012qj}) will allow observation of equilibration of the system including full transverse dynamics.

\section{Setup: Heavy-ion collisions as gravitational shock waves}

Boosting a charge to very high velocities, its energy-density
distribution becomes highly singular. This is very similar
to the case of boosting a mass to very high velocities.
However, in the latter case it has been understood by 
Aichelburg and Sexl in the 1970's that 
a reasonable description of the energy-momentum
tensor can be given by rescaling the mass with the boost factor
\cite{Aichelburg:1970dh}.
By complete analogy, the energy-momentum
tensor of a boosted charge $\rho$ can be calculated analytically
by means of a rescaling of the coupling constant \cite{Steinbauer:1996fv},
arriving at a form of
$$
T_{++}\propto \rho(x_\perp) \delta(x^+)\,,
$$
where light-cone coordinates $x^\pm=\frac{x^0\pm x^3}{\sqrt{2}}$ 
have been introduced and $x_\perp=(x^1,x^2)$ are the coordinates
in the plane transverse to the boosting direction.

Within gauge/gravity duality, it is known how to construct
a strongly coupled ${\cal N}=4$ SYM configuration that has precisely this
energy momentum tensor. The line element is given by
\begin{equation}
\label{origds2}
ds^2=\frac{-2 dx^+ dx^-+dx_\perp^2 + dz^2 + dx^{+ 2} \Phi(x_\perp,z)\delta(x^+)}{z^2}\,,
\end{equation}
where $z$ is the coordinate parametrizing the fifth dimension in 
a space that is asymptotically Anti-de-Sitter and 
\begin{equation}
\label{densitydef}
\lim_{z\rightarrow 0}\frac{\Phi(x_\perp,z)}{z^4}=\frac{\rho(x_\perp)}{\kappa}\,,
\end{equation}
where $\kappa$ is the number of degrees of freedom and we have taken $T_{++}=\rho(x_\perp)\delta(x^+)$.  Note that 
while for ${\cal N}=4$ SYM $\kappa=\frac{N_c^2}{2\pi^2}$,
it is easy to make the number of degrees of freedom more QCD-like
by resetting $\kappa$ by hand, as done e.g. in \cite{Wu:2011yd}.

This line element corresponds to a gravitational shock wave in $AdS_5$
with a transverse profile that is governed by the density
distribution $\rho(x_\perp)$. It is an exact solution to Einstein
equations
if \cite{Taliotis:2010pi}
\begin{equation}
\left[\partial_z^2-\frac{3}{z}\partial_z +\partial_\perp^2\right]
\Phi(x_\perp,z)=0\,,
\label{phieq}
\end{equation}
which can be solved in Fourier-space to give \cite{Avsar:2009xf}
$$
\Phi(k_\perp,z)=c_2(k_\perp) z^2 I_2(z k_\perp)\,,
$$
where $I_2$ is a modified Bessel function. Note that the 
second solution to (\ref{phieq}) does not describe a ${\cal N}=4$
SYM field theory in Minkowski space and therefore is not allowed here.
The function $c_2(k_\perp)$ is related to the Fourier-transform
of the profile function $\rho(k_\perp)$ by
$$
c_2(k_\perp)=\frac{8 \rho(k_\perp)}{\kappa k_\perp^2}\,.
$$

One should note that one could supplement the gauge/gravity
setup by introducing sources (currents) in the bulk, in order
to regulate the behavior of the profile function $\Phi(x_\perp,z)$
for large values of $z$.  However,
it turns out that for the purpose of shock wave collisions,
the results including these sources is identical to those without
sources \cite{Kovchegov:2009du}. Therefore, we chose to not modify the original 
gauge/gravity setup and work with profile functions that may
not have well defined $z\rightarrow \infty$ limits. Since
in the process of the shock wave collision, a horizon will
form at a finite value of $z$, the large $z$ behavior is
no longer relevant for the subsequent evolution anyways.

As a model for a boosted nucleus, the form of $\rho$ can
be taken to be a Wood-Saxon distribution, a distribution calculated
from the Color-Glass-Condensate model or any other favorable model.
The present technique for the pre-equilibrium dynamics
is not limited to any specific nuclear physics model for a boosted nucleus.

For example, one can take $\Phi(k_\perp,z)\propto z^2 I_2(z k_\perp) K_2(z_0 k_\perp)$,
which in configuration space can be shown to be
$$
\Phi(|x_\perp|,z)\propto z q^{-3}\  _2F_1(3,5/2,3,-1/q)\,,
$$
where $q=\frac{x_\perp^2+(z-z_0)^2}{4 z z_0}$ and $_2F_1$ is
a hypergeometric function that takes a particular simple
form \cite{Gubser:2008pc,Avsar:2009xf}. This particular
choice corresponds to a profile function 
$\rho(x_\perp)\propto (x_\perp^2+z_0^2)^{-3}$. For a recent 
study relating the form of $\rho(x_\perp)$ to the presence
(or absence) of trapped surfaces formed in the collision see
\cite{Taliotis:2012sx}.

\section{Methodology}

The first important point to address is generalizing
the matching conditions for the collision of two shock
waves with transverse profile $\rho={\rm const}$ in Ref.~\cite{Grumiller:2008va}
to arbitrary profiles $\rho(x_\perp)$.
To perform the matching, it is advisable to transform
the line element to a differentiable form that does not contain
a $\delta$-function (Rosen coordinates). For a general transverse
profile $\rho(x_\perp)$, this can be easily done by generalizing
the corresponding case for spaces that are asymptotically 
Minkowski \cite{Yoshino:2002br}. The relevant coordinate 
transformations for a single shock wave are
\begin{equation}
x^+=u\,,\quad
x^-=v+\frac{1}{2}\Phi \theta(u)+\frac{u \theta^2(u)}{8} \partial_i
\Phi\,\delta^{ij}\, \partial_j \Phi\,,\quad
x^i=\tilde x^i + \frac{1}{2} u \theta(u) \delta^{ij} \partial_j \Phi\,,
\end{equation}
where the coordinates $x^i=(x^1,x^2,z)$ were introduced. Superposing
the metrics from shock wave one and shock wave two one obtains a line
element that is valid before the collision:
\begin{equation}
\label{ds2pre}
ds^2_{\rm pre} = \frac{-2 du dv + d\tilde x^i d\tilde x^j \left(\delta^{kl}
H_{ik}^{(1)} H_{jl}^{(1)}+\delta^{kl} H_{ik}^{(2)} H_{jl}^{(2)}-\delta_{ij}\right)
}{\left[\tilde z+\frac{1}{2} \left(u \theta(u) \partial_z \Phi_{(1)} +
v \theta(v) \partial_z \Phi_{(2)}\right)\right]^2}\,,
\end{equation}
where $u<0,v<0$ and 
$$
H^{(1)}_{ij}=\delta_{ij} + \frac{u \theta(u)}{2}\partial_i \partial_j \Phi_{(1)}
\,,\quad
H^{(2)}_{ij}=\delta_{ij} + \frac{v \theta(v)}{2}\partial_i \partial_j \Phi_{(2)}\,.
$$

\paragraph{Central Collisions}

While the following program seems straightforward for any (physically allowed) pair of source functions $\Phi_{(1)},\Phi_{(2)}$, one can expect the calculation to be rather tedious. Therefore, in this article we limit ourselves to considering head-on collisions of azimuthally symmetric nuclei, for which azimuthal symmetry is unbroken in the future light cone. Hence one introduces new coordinates $r,\phi$
$$
\tilde x=r \cos\phi,\,,\quad \tilde y=r \sin\phi\,, \quad \tilde z=z
$$
for which one finds in particular\footnote{Note that the arguments of $\Phi$ in 
Eq.~(\ref{origds2}) can be take as $\tilde x^i$ because the $\delta$-function multiplying $\Phi$ erases the distinction between these. Clearly, the same argument will not hold true if one had had ``smeared'' delta functions instead.}
$$
\partial_i \partial_j \Phi = e_r^i e_r^j \partial_r^2 \Phi
+2 e_r^{(i} e_z^{j)} \partial_r \partial_z \Phi + e_\phi^i e_\phi^j \frac{1}{r}\partial_r \Phi+e_z^i e_z^{j} \partial_z^2 \Phi\,,
$$
because $\Phi=\Phi(r,z)$. For better readability, it is convenient to introduce the notation
$$
\Phi^{m,n} \equiv \partial_r^{m} \partial_z^{n} \Phi(r,z)\,.
$$

The pre-collision line element $ds_{\rm pre}$ then becomes
$$
ds_{\rm pre}^2=\frac{-2 du dv + d\tilde x^i d\tilde x^j M_{ij}^{\rm pre}
}{\left[\tilde z+\frac{1}{2} \left(u \theta(u) \Phi_{(1)}^{0,1} +
v \theta(v)\Phi_{(2)}^{0,1}\right)\right]^2}\,,
$$
with
\begin{eqnarray}
\label{metricfuncs1}
M^{pre}_{rr}&=&1+u \theta(u) \Phi^{2,0}_{(1)} +\frac{u^2 \theta^2(u)}{4} \left(\left(\Phi^{2,0}_{(1)}\right)^2
+\left(\Phi^{1,1}_{(1)}\right)^2\right)\nonumber\\
&&\hspace*{0.5cm} +v \theta(v) \Phi^{2,0}_{(2)}
+\frac{v^2 \theta^2(v)}{4}\left(\left(\Phi^{2,0}_{(2)}\right)^2
+\left(\Phi^{1,1}_{(2)}\right)^2\right)\,,
\nonumber\\
M^{pre}_{\phi\phi}&=&1+ \frac{u \theta(u)}{r} \Phi^{1,0}_{(1)}+\frac{u^2 \theta^2(u)}{4 r^2} \left(\Phi^{1,0}_{(1)}\right)^2 
+ \left[u\leftrightarrow v, 1\leftrightarrow 2\right]\,,
\nonumber\\
M^{pre}_{zz}&=&1+u \theta(u) \Phi^{0,2}_{(1)} + \frac{u^2 \theta^2(u)}{4}
 \left(\left(\Phi^{0,2}_{(1)}\right)^2
+\left(\Phi^{1,1}_{(1)}\right)^2\right)
+ \left[u\leftrightarrow v, 1\leftrightarrow 2\right]\,,
\nonumber\\
M^{pre}_{rz}&=&u \theta(u) \Phi^{1,1}_{(1)}+\frac{u^2 \theta(u)^2}{4} \Phi^{1,1}_{(1)}\left(\Phi^{2,0}_{(1)}+\Phi^{0,2}_{(1)}\right)
+ \left[u\leftrightarrow v, 1\leftrightarrow 2\right]\,,
\end{eqnarray}
and all others vanishing.

\paragraph{Matching}
The matching at the collision point is performed by making an ansatz
for the line element such as
\begin{equation}
\label{lineel}
ds^2=ds_{\rm pre}^2+\theta(u)\theta(v)ds_{\rm int}^2
\end{equation}
with $ds_{\rm pre}^2$ given in Eq.~(\ref{ds2pre}). The number of independent metric functions appearing in the interaction part of the line element $ds_{\rm int}^2$ can be found be with the following argument. First note that if the line element $ds^2$ was in Fefferman-Graham form, then this would fix the metric function $g_{zz}$ to be $1/z^2$ and $g_{u z},g_{v z},g_{r z}, g_{\phi z}$ to vanish. The non-vanishing components of the metric would be in a sub-matrix spanned by the coordinates $u,v,r,\phi$, so in principle there would be $10$ independent components. Limiting ourselves to systems that have azimuthal symmetry implies that $g_{u\phi},g_{v \phi},g_{r \phi}$ are vanishing, and hence one only has $7$ independent metric functions. For \emph{symmetric} collisions (e.g. $\Phi_{(1)}=\Phi_{(2)}$), one may thus pose
\begin{eqnarray}
\label{ds2post}
ds^2_{\rm int}&=&-2 du dv M_{u v}+\left(\frac{u}{v} dv^2+\frac{v}{u} du^2\right) M_{uu}\nonumber\\
&&+2 dr (du+dv) M_{u r}+ dr^2 M_{rr}+d\phi^2 M_{\phi\phi}+dz^2 M_{zz}+2 dr dz M_{rz}\,.
\end{eqnarray}
Asymmetric central collisions (to model for instance proton-nucleus (pA) collisions) will be treated separately below. Expanding $M_{ab}$ around $u=0,v=0$, we demand that the Einstein equations
are fulfilled across the the light-cone. 
In essence, this is equivalent to the procedure 
performed for the so-called Color Glass Condensate~\cite{Kovner:1995ja}, 
where one deals
with Yang-Mills equations instead of Einstein equations and one matches
gauge field configurations instead of the metric field. 

Close to the light-cone $u\simeq 0,v\simeq 0$, so one can make an ansatz such as
\begin{eqnarray}
\label{ansatz}
M_{rr}(u,v,r,z)&=&(u+v) f_{rr}^{11}(r,z)+ u v f_{rr}^{20}(r,z) + (u^2+v^2)f_{rr}^{22}(r,z)\nonumber\\
&&+u v (u+v) f_{rr}^{31}(r,z)+(u^3+v^3)f_{rr}^{33}(r,z)+u^2 v^2 f_{rr}^{40}(r,z)
\nonumber\\
&&+u v (u^2+v^2)f_{rr}^{42}(r,z)+(u^4+v^4)f_{rr}^{44}(r,z)+\ldots\,,
\end{eqnarray}
and equivalently for the other metric functions $M_{ab}$.
Note that absence of negative powers of $u,v$ simply follows from the observation that these would give rise to highly singular terms such as $\delta'(u)/u$ in the Einstein Equations. 

In order to solve Einstein's equations on the light-cone ($u=0$ or $v=0$), the coefficient for all singular functions has to vanish. Specifically, one finds that the condition that there are no $\delta(u)/u$, $\delta^2(u)$ or $\delta'(u)$ terms in the $u,v$ component of Einstein's equations immediately leads to
$$
f_{ij}^{11}=0\,,f^{22}_{ij}=0\,,f^{33}_{ij}=0\,,f^{44}_{ij}=0\,\ldots
$$
meaning that the line element has to be \emph{continuous} across the light cone $u=0$, $v=0$. Using this information, it is vastly more convenient 
to again change coordinates using the so-called Milne form
$$
\tau = \sqrt{2 u v}\,,\quad \xi = \frac{1}{2}\ln\frac{u}{v}\,,
$$
where
\begin{equation}
\label{txds2}
ds^2=\frac{-d\tau^2 g_{\tau\tau}+\tau^2 d\xi^2 g_{\xi\xi}+dr^2 g_{rr}+r^2 d\phi^2 g_{\phi\phi}+2 dr dz g_{rz}+2 \tau d\tau dr g_{\tau r}+ dz^2 g_{zz}}
{\left[z + \frac{\tau}{\sqrt{2}} \left(e^{\xi}\Phi^{0,1}_{(1)}+e^{-\xi} \Phi^{0,1}_{(2)}\right)\right]^2}
\,,
\end{equation}
and metric functions
\begin{eqnarray}
g_{\tau \tau}&=&1+K(\tau,\xi,r,z)\,,\nonumber\\
g_{\xi\xi}&=&1+L(\tau,\xi,r,z)\,,\nonumber\\
g_{rr}&=&M_{rr}^{pre}\left(1+H(\tau,\xi,r,z)\right)\,,\nonumber\\
g_{\phi\phi}&=&M_{\phi\phi}^{pre}\left(1+F(\tau,\xi,r,z)\right)\,,\nonumber\\
g_{rz}&=&M_{rz}^{pre}+G(\tau,\xi,r,z)\,,\nonumber\\
g_{\tau r}&=&\tau J(\tau,\xi,r,z)\,,\nonumber\\
g_{zz}&=&M_{zz}^{pre}\left(1+M(\tau,\xi,r,z)\right)\,.\nonumber\\
\end{eqnarray}
Since this is a central point of our work, let us stress that we do not assume the metric to be continuous across the light-cone. In fact, Eq.~(\ref{ansatz}) does contain terms that imply the metric to jump at the light-cone. However, when solving Einstein's equations, we find that the coefficients of these terms have to vanish, otherwise there is no solution. The continuity of the metric across the light-cone is a result, not an assumption, of our work.

For early times $\tau\ll 1$, the seven metric coefficient functions $J,K,L,H,F,G,M$ may be expanded in a Taylor series around $\tau=0$ and the coefficients of this Taylor series are determined by solving the Einstein Equations order by order in $\tau$. The resulting line element $ds^2$ may then be brought into a more convenient form, such as Fefferman-Graham coordinates or Eddington-Finkelstein coordinates by a suitable coordinate transformation. 

\section{Solution for central AA collisions}
\label{sec:AA}

Using the methodology outlined in the last section one can find a solution the case of a head-on collision of two shock waves with \emph{identical} profile functions, $\Phi_{(1)}=\Phi_{(2)}\equiv\Phi$. This could be interpreted as a model for nucleus-nucleus (AA) collisions, such as Pb-Pb at LHC energies. In this case, the metric functions $J,K,L,H,F,G,M$ can be expanded as
$$
K=\tau^2 k_{20}(r,z)+\tau^3 \cosh\xi\, k_{31}(r,z)+\tau^4 k_{40}(r,z)+\tau^4 \cosh{2 \xi}\, k_{42}(r,z)+\ldots\,,
$$
which corresponds to the expansion in Eq.~(\ref{ansatz}) plus the additional knowledge that $K$ must vanish for $u=0$ or $v=0$ (absence of singularities on the light-cone). 

Solving the Einstein Equations order by order in $\tau$ one finds for example
\begin{eqnarray}
l_{20}&=&
\frac{k_{20}}{3}-\frac{2 \left(\Phi^{0,1}\right)^2}{3 z^2}+\frac{\Phi^{0,1} \left(\Phi^{1,0}-r \Phi^{2,0}\right)}{2 r z}-\frac{\left(\Phi^{1,0}\right)^2}{6 r^2}-\frac{\Phi^{1,0} \Phi^{2,0}}{6 r}-\frac{\left(\Phi^{1,1}\right)^2+\left(\Phi^{2,0}\right)^2}{6}\,,\nonumber\\
h_{20}&=&-\frac{\left(\Phi^{0,1}\right)^2}{4 z^2}+\frac{\left(\Phi^{1,1}\right)^2+\left(\Phi^{2,0}\right)^2}{2}\,,\nonumber\\
f_{20}&=&-\frac{\left(\Phi^{0,1}\right)^2}{4 z^2}+\frac{\left(\Phi^{1,0}\right)^2}{2 r^2}\,,\\
g_{20}&=&\frac{3 \Phi^{0,1}}{2 z}-\frac{\Phi^{1,0}\Phi^{1,1}}{2 r}\,,\nonumber\\
m_{20}&=&\frac{17 \left(\Phi^{0,1}\right)^2}{4 z^2}-\frac{3 \Phi^{0,1}\Phi^{1,0}}{r z}+\frac{\left(\Phi^{1,0}\right)^2}{2 r^2}+\frac{\left(\Phi^{1,1}\right)^2}{2}
-\frac{3 \Phi^{0,1}\Phi^{2,0}}{z}+\frac{\Phi^{1,0}\Phi^{2,0}}{r}
+\frac{\left(\Phi^{2,0}\right)^2}{2}\,,\nonumber
\end{eqnarray}
while the expression for $j_{20}$ is too lengthy to be reproduced here and is presented in appendix \ref{app: j20}.
Note that the coefficient function $K$ is only determined as a constraint at higher orders. Specifically, one find that its early-time, near-boundary expansion\footnote{Note that while the expansion is formally in $\tau$, the actual terms in appearing in the series are of the form $\tau z^2 \cosh\xi$. As a consequence, for any non-zero $\tau$, we do not expect the series to converge except for central-rapidity $\xi\simeq 0$ and close to the boundary $z\simeq 0$.} is given by
$$
k_{20}=-\frac{5\left(\left.\Phi^{0,4}\right|_{z=0}\right)^2 z^4}{288}+{\cal O}(z^6)=-\frac{10}{\kappa^2} \rho^2(r) z^4+{\cal O}(z^6)\,,
$$
where we expressed the result in terms of the charge density $\rho$ of the original shock wave (nucleus) and degrees of freedom $\kappa$ (cf. Eq.~\ref{densitydef}).

With the post-collision metric known, one would like to extract information about the boundary energy-momentum tensor $T^{\mu\nu}$ where $\mu=0,1,2,3$ and $x^{\mu}=(\tau,x^1,x^2,\xi)$. This is most easily achieved by rewriting the line element in Fefferman-Graham form\,,
\begin{equation}
\label{ds2FG}
ds^2=\frac{g_{\mu\nu}dx^\mu dx^\nu+dz^2}{z^2}+\frac{z^4 T_{\mu\nu} dx^\mu dx^\nu}{\kappa z^2}+\sum_{n=0}^\infty \frac{z^{6+2 n}h_{\mu\nu}^{(n)} dx^\mu dx^\nu}{z^2}\,,
\end{equation}
where the boundary metric is assumed to be flat: $g_{\mu\nu}={\rm diag}(-1,1,1,\tau^2)$. To bring the line element Eq.~(\ref{txds2}) into the Fefferman-Graham form, one uses the coordinate transformation
\begin{eqnarray}
\label{cootra}
\tau&=&\tau_{FG}+\sum_{n=0}^\infty t_n(\tau_{FG},\xi_{FG},r_{FG}) z^{4+2n}_{FG}\,,\nonumber\\
\xi&=&\xi_{FG}+\sum_{n=0}^\infty e_n(\tau_{FG},\xi_{FG},r_{FG}) z^{4+2n}_{FG}\,,\nonumber\\
r&=&r_{FG}+\sum_{n=0}^\infty s_n(\tau_{FG},\xi_{FG},r_{FG}) z^{4+2n}_{FG}\,,\nonumber\\
z&=&z_{FG}+\sum_{n=0}^\infty a_n(\tau_{FG},\xi_{FG},r_{FG}) z^{3+2n}_{FG}\,,\nonumber\\
\phi&=&\phi_{FG}\,,
\end{eqnarray}
with coefficients $t_n,e_n,s_n,a_n$ which are determined order by order in $z$.
One finds that to lowest order the required coordinate transformations are
\begin{eqnarray}
t_0&=&-\frac{\cosh\xi_{FG} \rho(r_{RG})}{\sqrt{2}\kappa}+{\cal O}(\tau^4)\,,\nonumber\\
e_0&=&\frac{\sinh\xi_{FG} \rho(r_{RG})}{\sqrt{2}\tau \kappa}+{\cal O}(\tau^3)\,,\nonumber\\
s_0&=&-\frac{\tau_{FG}\cosh\xi_{FG}  \rho^\prime(r_{RG})}{\sqrt{2}\kappa}+{\cal O}(\tau^5)\,,\nonumber\\
a_0&=&-4 \frac{\tau_{FG} \cosh\xi_{FG}\rho(r_{RG}) }{ \sqrt{2}\kappa}+{\cal O}(\tau^5)\,.
\end{eqnarray}
Since at $z=z_{FG}=0$ we have $\tau_{FG}=\tau$, the resulting boundary energy-momentum tensor may be written in terms of the boundary coordinates $x^\mu$. Solving Einstein's equations up to (including) ${\cal O}(\tau^3)$ one finds
\begin{eqnarray}
\label{emt}
T_{\tau\tau}&=&\frac{2}{\kappa} \rho(r)^2\tau^2+ {\cal O}(\tau^4)\,,
\nonumber\\
\tau^{-2}T_{\xi\xi}&=&-\frac{6}{\kappa}\rho(r)^2 \tau^2 + {\cal O}(\tau^4)\,,\nonumber\\
T_{rr}&=&\frac{4}{\kappa} \rho(r)^2\tau^2+\left(6 \rho^{\prime 2}(r)-\frac{\rho(r) \rho^\prime(r)}{r}-3 \rho(r)\rho^{\prime\prime}(r)\right)\frac{\tau^4}{2 \kappa}+{\cal O}(\tau^5)\,,\nonumber\\
T_{\phi\phi}&=&4\frac{\rho^2(r)}{\kappa} r^2 \tau^2+
\left(2\rho^{\prime 2}(r)-3 \frac{\rho(r)\rho^\prime(r)}{r}- \rho(r)\rho^{\prime\prime}(r)\right)\frac{r^2 \tau^4}{2\kappa}+{\cal O}(\tau^5)\,,\nonumber\\
T_{r \tau}&=&2\frac{\rho(r)\rho^\prime(r)\tau^3}{\kappa}+{\cal O}(\tau^5)\,,\nonumber\\
T_{\xi \tau}&=&{\cal O}(\tau^5)\,,\nonumber\\
T_{\xi r}&=&{\cal O}(\tau^5)\,.
\end{eqnarray}
Note that because of the nature of the expansion, this result is expect to hold only close to mid-rapidity $\xi\simeq 0$. Up to the order given, this energy-momentum tensor is traceless and covariantly conserved, $\nabla_\mu T^{\mu\nu}=0$. In fluid dynamics, it is custom to decompose the energy-momentum tensor using the energy density $\epsilon$ and the fluid velocity $u^\mu$ with $u_\mu u^\mu=-1$. Even for non-equilibrium (dissipative) fluids, these are completely defined in terms of the eigenvalues and eigenvectors of the energy momentum tensor
$$ 
u_\mu T^{\mu\nu}=-\epsilon u^\nu\,.
$$
Ignoring for the moment the fact that the solution Eq.~(\ref{emt}) is not that of a fluid, one may still ask what values of $\epsilon$ and $u^\mu$ the energy-momentum tensor corresponds to if one \emph{pretended} it was that of a fluid. One finds
\begin{equation}
\epsilon=\frac{2}{\kappa} \rho^2(r) \tau^2\,,\quad
u^r=-\frac{\rho^\prime(r)}{3 \rho(r)}\tau+{\cal O}(\tau^3)\,, \quad u^\xi={\cal O}(\tau)\,.
\end{equation}
This finding implies that there is a radial flow $u^r$ building up that is proportional to the gradient of the transverse charge distribution of the nucleus,
\begin{equation}
\partial_\tau u^r=-\frac{\partial_r \ln(\rho(r))}{3}\,,
\end{equation}
which should be compared to the evolution expected for \emph{ideal} hydrodynamics with conformal equation of state ($c_s^2=1/3$) \cite{Baier:2006gy}:
\begin{equation}
\label{idealhydro}
\partial_\tau u^r_{\rm ideal\, hydro} = -c_s^2 \partial_r \ln s\,,
\end{equation}
where $s$ is the initial entropy density. Clearly, the evolution found for here for the early-time pre-equilibrium radial flow is \emph{identical} to that from linear ideal hydrodynamics, provided one identifies the square root of the initial overlap distribution $\sqrt{\rho^2(r)}$ with the entropy density $s(r)$. The same is not true for the extracted energy density $\epsilon$, which according to linear hydrodynamics should decrease rather than increase. In any case, the present calculation provides a concrete example for a far-from equilibrium system evolution with a flow profile identical to that expected from ideal hydrodynamics. If a similar phenomenon were to happen for anisotropic flow, this would have important implications for the attempt to use experimental anisotropic flow measurements to extract the viscosity coefficient of hot QCD matter.

\section{Solution for central pA collisions}

The above strategy to find the line element after the collision of two shock waves may be generalized to the case of asymmetric collisions ($\Phi_1\neq\Phi_2$), which may be taken to represent a model for the proton-nucleus collisions (pA). In this case, there is no longer a symmetry $u\leftrightarrow v$ and hence it is advisable to slightly generalize the ansatz for the post-collision line element $ds_{\rm post}$ (\ref{ds2post}). Matching predicts that the line element is again continuous across the light-cone, so one may switch again to Milne coordinates $\tau,\xi$. We find that the line element
\begin{eqnarray}
\label{ds2pA}
ds^2&=&
\left[z + \frac{\tau}{\sqrt{2}} \left(e^{\xi}\Phi^{0,1}_{(1)}+e^{-\xi} \Phi^{0,1}_{(2)}\right)\right]^{-2}
\times
\left[-d\tau^2 g_{\tau\tau}+\tau^2 d\xi (g_{\xi\xi} d\xi+2 g_{\xi r}dr+2 g_{\xi z} dz)\right.\nonumber\\
&&\left.
+
dr (g_{rr}d r+2 g_{rz} dz+2 \tau g_{\tau r} d\tau)+r^2 d\phi^2 g_{\phi\phi}+ dz^2 g_{zz}\right]
\,,
\end{eqnarray}
is suitable for obtaining a solution to the Einstein Equations, even though (or maybe because) it contains two redundant metric functions. Following the strategy outlined in section \ref{sec:AA}, one can solve the Einstein Equations order by order in a power series expansion in $\tau$. The resulting solution may then be transformed to Fefferman-Graham coordinates and one finds the following result for the energy-momentum tensor:
\begin{eqnarray}
\label{emtpA}
T_{\tau\tau}&=&\frac{2}{\kappa} \rho_1(r) \rho_2(r)\tau^2+ {\cal O}(\tau^4)\,,
\nonumber\\
\tau^{-2}T_{\xi\xi}&=&-\frac{6}{\kappa}\rho_1(r) \rho_2(r) \tau^2 + {\cal O}(\tau^4)\,,\nonumber\\
T_{rr}&=&\frac{4}{\kappa} \rho_1(r)\rho_2(r)\tau^2+{\cal O}(\tau^4)\,,\nonumber\\
T_{\phi\phi}&=&4\frac{\rho_1(r) \rho_2(r)}{\kappa} r^2 \tau^2+{\cal O}(\tau^4)\,,\nonumber\\
T_{r \tau}&=&\frac{\rho_1^\prime(r) \rho_2(r)+\rho_1(r)\rho_2^\prime(r)}{\kappa}\tau^3+{\cal O}(\tau^5)\,,\nonumber\\
T_{\xi \tau}&=&{\cal O}(\tau^5)\,,\nonumber\\
T_{\xi r}&=&3 \frac{\rho_1^\prime(r)\rho_2(r)-\rho_1\rho_2^\prime(r)}{\kappa}\tau^4{\cal O}(\tau^5)\,.
\end{eqnarray}
As it should, this result corresponds to the form for AA collisions when $\Phi_1=\Phi_2$. Note that in this case one obtains a pre-equilibrium radial flow
\begin{equation}
\label{urpA}
u^r=-\frac{\rho_1^\prime(r)\rho_2(r)+\rho_1(r)\rho_2^\prime(r)}{6 \rho_1(r)\rho_2(r)}\,,
%
\end{equation}
which again can be interpreted as \emph{ideal} hydrodynamic flow buildup (\ref{idealhydro}) for an entropy density
$$
s(r)\propto 
\sqrt{\rho_1(r)\rho_2(r)}\,,
$$
with $\rho_{1,2}(r)$ the transverse charge density distribution of nucleus $1,2$, respectively.

\section{Summary and Conclusions}

In this article, we studied the central collision of gravitational shock waves in asymptotic $AdS_5$ spacetimes.
For early times after the collision, we find an explicit, systematically improvable result for the post-collision line element in form of a series expansion that is well-behaved at mid-rapidity and close to the AdS boundary. 
The novel aspect about our study is that we allowed the shock waves to have arbitrary profiles with azimuthal symmetry in the plane transverse to the collision axis, thus generalizing the result of Ref.~\cite{Grumiller:2008va}. Via gauge/gravity duality, we are able to interpret our result as the early-time energy momentum tensor generated by the collision of two "nuclei" in ${\cal N}=4$ SYM. 

Our three most important findings are: 
\begin{enumerate}
\item
It is possible to generalize the methods developed in cf.~\cite{Grumiller:2008va} to the case of incident shock waves with less symmetry, thereby indicating that the technique should also be applicable to gravitational shock waves with no special symmetries in the transverse plane. 
\item
The resulting early-time energy-momentum tensor reflects the fact that the system is
initially far from equilibrium. For instance, the effective longitudinal pressure is negative.
Our result for the pre-equilibrium energy-momentum tensor implies matter flow in the (transverse) radial direction. This serves as a concrete example for the generation of far-from equilibrium flow in high energy "nuclear" collisions. 
\item
The build-up of this radial flow is identical to that expected from ideal hydrodynamics with an entropy density proportional to the square root of the product of the charge density of the individual shock waves. This last result is somewhat unexpected because the system, being far from equilibrium, does not evolve according to ideal hydrodynamics as a whole; only the radial flow buildup does. 
\end{enumerate}

We also studied the implications for non-symmetric collisions as a model of proton-nucleus (pA) collisions. In this case, all three of the above points also apply. In particular, our result implies strong (as compared to AA collisions) early-time radial flow as a consequence of the proton charge density falling of much more steeply in the radial direction than that for a heavy nucleus. This could possibly explain some of the experimental findings in proton-lead and proton-gold collisions at high energies.

The biggest limitation of the present study is that it is only applicable at very early times where a power series expansion converges. However, there is a good chance that these early time results may be used as input for a subsequent numerical solution of the Einstein Equations (e.g. from Ref.~\cite{vanderSchee:2012qj}) to obtain full results for all subsequent times, as in Ref.~\cite{Wu:2011yd}. We intend to pursue this direction in a future publication. 

As another application of the present work, we want to emphasize that there seems to be no obstacle in generalizing the present results to the case of shock wave collisions with arbitrary transverse profiles $\rho(x_\perp)$, such as provided by the PHOBOS Monte-Carlo Glauber model \cite{Alver:2008aq}, which results in a lumpy distribution for $\rho$.
In this case, subsequent results for the post-collision dynamics would have to be averaged over many configurations (events), yet from the experience gained in the present work, also this case seems to be feasible using known techniques. 

\section*{Acknowledgments
}

This work was supported by the Department of Energy, grant award No. DE-SC0008132. PR would like to thank A.~ Taliotis and W.~van der Schee for fruitful discussions.


\appendix
\section{J20}
\label{app: j20}
\begin{eqnarray}
\label{j20}
j_{20}&=&\Bigg[48 r^3 z^4 k^{0,1}_{31} + \sqrt{2} \bigg( -546 r^3 \left(\Phi^{0,1} \right)^3 + 687 r^2 z \left(\Phi^{0,1}\right)^2 \Phi^{1,0}\nonumber\\
&& - 287 r z^2 \Phi^{0,1}\left(\Phi^{1,0}\right)^2 + 40 z^3 \left(\Phi^{1,0}\right)^3 + 120 r^2 z^2 \left( \Phi^{0,1} \right)^2 \Phi^{1,1} - 24 r^3 z^4 k^{1,0}_{20} \Phi^{1,1}\nonumber\\
&& -250 r z^3 \Phi^{0,1}\Phi^{1,0}\Phi^{1,1} + 90 z^4 \left(\Phi^{1,0}\right)^2\Phi^{1,1} - 764 r^3 z ^2 \Phi^{0,1} \left(\Phi^{1,1}\right)^2\nonumber\\
&&  +72 r^3 z^4 \Phi^{0,3}\left(\Phi^{1,1}\right)^2 + 346 r^2 z^3 \Phi^{1,0} \left(\Phi^{1,1}\right)^2 + 72 r^2 z^4 \left(\Phi^{1,1}\right)^3 \nonumber\\
&&  +687 r^3 z \left(\Phi^{0,1}\right)^2 \Phi^{2,0} - 584 r^2 z^2 \Phi^{0,1} \Phi^{1,0} \Phi^{2,0} + 125 r z^3 \left(\Phi^{1,0}\right)^2 \Phi^{2,0}  \nonumber\\
&&  + 40 r^2 z^3 \Phi^{0,1} \Phi^{1,1} \Phi^{2,0} + 30 r z^4 \Phi^{1,0} \Phi^{1,1} \Phi^{2,0} + 346 r^3 z^3 \left(\Phi^{1,1}\right)^2 \Phi^{2,0} \nonumber\\
&&  - 287 r^3 z^2 \Phi^{0,1} \left(\Phi^{2,0}\right)^2 + 125 r^2 z^3 \Phi^{1,0} \left(\Phi^{2,0}\right)^2 - 30 r^2 z^4 \Phi^{1,1} \left(\Phi^{2,0}\right)^2 \nonumber\\
&& + 40 r^3 z^3 \left(\Phi^{2,0}\right)^3 - 8 r^2 z^2 k_{20} \left(-2 r \Phi^{0,1} + z \left(\Phi^{1,0} + r \Phi^{2,0}\right)\right) \nonumber\\
&&  + 8 r^2 z^3 k^{0,1}_{20} \left(-8 r \Phi^{0,1} + 3 z \left( \Phi^{1,0} + r \Phi^{2,0}\right)\right) + 120 r^3 z^2 \left(\Phi^{0,1}\right)^2 \Phi^{2,1} \nonumber\\
&&  - 85 r^2 z^3 \Phi^{0,1} \Phi^{1,0} \Phi^{2,1} + 15 r z^4 \left(\Phi^{1,0}\right)^2 \Phi^{2,1} + 42 r^3 z^4 \left(\Phi^{1,1}\right)^2 \Phi^{2,1} \nonumber \\
&& - 125 r^3 z^3 \Phi^{0,1} \Phi^{2,0} \Phi^{2,1} + 45 r^2 z^4 \Phi^{1,0} \Phi^{2,0} \Phi^{2,1} + 30 r^3 z^4 \left( \Phi^{2,0} \right)^2 \Phi^{2,1} \nonumber\\
&& + 125 r^3 z^3 \Phi^{0,1} \Phi^{1,1} \Phi^{3,0} - 45 r^2 z^4 \Phi^{1,0} \Phi^{1,1} \Phi^{3,0} \nonumber\\
&& - 60 r^3 z^4 \Phi^{1,1} \Phi^{2,0} \Phi^{3,0} \bigg) \Bigg] \Bigg/ \left( 144 \sqrt{2} r^3 z^4 \Phi^{1,1}\right)\,,
\end{eqnarray}
where the coefficient function $k_{31}(r,z)$ is only known in the near boundary expansion. It's first non-vanishing term in the near-boundary expansion is related to $k_{20}$ and $\rho$ as
\begin{equation}
k_{31}(r,z)=\frac{480 \kappa^2 \rho(r) k_{20}^{0,4}(r,0)+728 \rho^3(r)}{3 \sqrt{2} \kappa^3}z^6+{\cal O}(z^8)\,.
\end{equation}

For convenience, the full calculation including all the metric coefficients can be found at \cite{website}.

\bibliographystyle{plain}

\begin{thebibliography}{99}

\bibitem{Adcox:2004mh} 
  K.~Adcox {\it et al.}  [PHENIX Collaboration],
  Nucl.\ Phys.\  A {\bf 757}, 184 (2005).

\bibitem{Back:2004je}
  B.~B.~Back {\it et al.},
  Nucl.\ Phys.\  A {\bf 757}, 28 (2005).

\bibitem{Arsene:2004fa}
  I.~Arsene {\it et al.}  [BRAHMS Collaboration],
  Nucl.\ Phys.\  A {\bf 757}, 1 (2005).

\bibitem{Adams:2005dq}
  J.~Adams {\it et al.}  [STAR Collaboration],
  Nucl.\ Phys.\  A {\bf 757}, 102 (2005).

\bibitem{Aamodt:2010pa}
  KAamodt {\it et al.}  [ALICE Collaboration],
  Phys.\ Rev.\ Lett.\  {\bf 105} (2010) 252302
  [arXiv:1011.3914 [nucl-ex]].

\bibitem{Aad:2010bu}
  G.~Aad {\it et al.}  [Atlas Collaboration],
  Phys.\ Rev.\ Lett.\  {\bf 105} (2010) 252303
  [arXiv:1011.6182 [hep-ex]].

\bibitem{Chatrchyan:2011sx}
  S.~Chatrchyan {\it et al.}  [CMS Collaboration],
  Phys.\ Rev.\ C {\bf 84} (2011) 024906
  [arXiv:1102.1957 [nucl-ex]].

\bibitem{Baier:2000sb}
  R.~Baier, A.~H.~Mueller, D.~Schiff and D.~T.~Son,
  Phys.\ Lett.\ B {\bf 502} (2001) 51
  [hep-ph/0009237].

\bibitem{Bjoraker:2000cf}
  J.~Bjoraker and R.~Venugopalan,
  Phys.\ Rev.\ C {\bf 63} (2001) 024609
  [hep-ph/0008294].


\bibitem{Arnold:2004ti}
  P.~B.~Arnold, J.~Lenaghan, G.~D.~Moore and L.~G.~Yaffe,
  Phys.\ Rev.\ Lett.\  {\bf 94} (2005) 072302
  [nucl-th/0409068].


\bibitem{Xu:2004mz}
  Z.~Xu and C.~Greiner,
  Phys.\ Rev.\ C {\bf 71} (2005) 064901
  [hep-ph/0406278].

\bibitem{Rebhan:2004ur}
  A.~Rebhan, P.~Romatschke and M.~Strickland,
  Phys.\ Rev.\ Lett.\  {\bf 94} (2005) 102303
  [hep-ph/0412016].

\bibitem{Romatschke:2005pm}
  P.~Romatschke and R.~Venugopalan,
  Phys.\ Rev.\ Lett.\  {\bf 96} (2006) 062302
  [hep-ph/0510121].

\bibitem{Berges:2008zt}
  J.~Berges, D.~Gelfand, S.~Scheffler and D.~Sexty,
  Phys.\ Lett.\ B {\bf 677} (2009) 210
  [arXiv:0812.3859 [hep-ph]].


\bibitem{Kurkela:2011ub}
  A.~Kurkela and G.~D.~Moore,
  JHEP {\bf 1111} (2011) 120
  [arXiv:1108.4684 [hep-ph]].


\bibitem{Dusling:2012ig}
  K.~Dusling, T.~Epelbaum, F.~Gelis and R.~Venugopalan,
  Phys.\ Rev.\ D {\bf 86} (2012) 085040
  [arXiv:1206.3336 [hep-ph]].

\bibitem{Janik:2005zt}
  R.~A.~Janik and R.~B.~Peschanski,
  Phys.\ Rev.\ D {\bf 73} (2006) 045013
  [hep-th/0512162].

\bibitem{Lin:2006rf}
  S.~Lin and E.~Shuryak,
  Phys.\ Rev.\ D {\bf 77} (2008) 085013
  [hep-ph/0610168].


\bibitem{Kovchegov:2007pq}
  Y.~V.~Kovchegov and A.~Taliotis,
  Phys.\ Rev.\ C {\bf 76} (2007) 014905
  [arXiv:0705.1234 [hep-ph]].

\bibitem{Gubser:2008pc}
  S.~S.~Gubser, S.~S.~Pufu and A.~Yarom,
  Phys.\ Rev.\ D {\bf 78} (2008) 066014
  [arXiv:0805.1551 [hep-th]].




\bibitem{Albacete:2008vs}
  J.~L.~Albacete, Y.~V.~Kovchegov and A.~Taliotis,
  JHEP {\bf 0807} (2008) 100
  [arXiv:0805.2927 [hep-th]].

\bibitem{Grumiller:2008va}
  D.~Grumiller, P.~Romatschke,
  JHEP {\bf 0808 } (2008)  027.
  [arXiv:0803.3226 [hep-th]].

\bibitem{AlvarezGaume:2008fx}
  L.~Alvarez-Gaume, C.~Gomez, A.~Sabio Vera, A.~Tavanfar and M.~A.~Vazquez-Mozo,
  JHEP {\bf 0902} (2009) 009
  [arXiv:0811.3969 [hep-th]].


\bibitem{Chesler:2008hg}
  P.~M.~Chesler, L.~G.~Yaffe,
  Phys.\ Rev.\ Lett.\  {\bf 102 } (2009)  211601.
  [arXiv:0812.2053 [hep-th]].

\bibitem{Chesler:2010bi}
  P.~M.~Chesler, L.~G.~Yaffe,
  Phys.\ Rev.\ Lett.\  {\bf 106 } (2011)  021601.
  [arXiv:1011.3562 [hep-th]].


\bibitem{Wu:2011yd}
  B.~Wu, P.~Romatschke,
in press, International Journal of Modern Physics C, 
DOI No: 10.1142/S0129183111016920, [arXiv:1108.3715 [hep-th]].

\bibitem{Balasubramanian:2011ur}
  V.~Balasubramanian, A.~Bernamonti, J.~de Boer, N.~Copland, B.~Craps, E.~Keski-Vakkuri, B.~Muller and A.~Schafer {\it et al.},
  Phys.\ Rev.\ D {\bf 84} (2011) 026010
  [arXiv:1103.2683 [hep-th]].


\bibitem{Mateos:2011tv}
  D.~Mateos and D.~Trancanelli,
  JHEP {\bf 1107} (2011) 054
  [arXiv:1106.1637 [hep-th]].

\bibitem{Kiritsis:2011yn}
  E.~Kiritsis and A.~Taliotis,
  JHEP {\bf 1204} (2012) 065
  [arXiv:1111.1931 [hep-ph]].

\bibitem{Mrowczynski:1993qm}
  S.~Mrowczynski,
  Phys.\ Lett.\ B {\bf 314} (1993) 118.

\bibitem{Son:2007vk}
  D.~T.~Son and A.~O.~Starinets,
  Ann.\ Rev.\ Nucl.\ Part.\ Sci.\  {\bf 57} (2007) 95
  [arXiv:0704.0240 [hep-th]].


\bibitem{vanderSchee:2012qj}
  W.~van der Schee,
  arXiv:1211.2218 [hep-th].




\bibitem{Aichelburg:1970dh}
  P.~C.~Aichelburg, R.~U.~Sexl,
  Gen.\ Rel.\ Grav.\  {\bf 2 } (1971)  303-312.
  


\bibitem{Steinbauer:1996fv}
  R.~Steinbauer,
  J.\ Math.\ Phys.\  {\bf 38 } (1997)  1614-1622.
  [gr-qc/9606059].






\bibitem{Taliotis:2010pi}
  A.~Taliotis,
  JHEP {\bf 1009 } (2010)  102.
  [arXiv:1004.3500 [hep-th]].

\bibitem{Avsar:2009xf}
  E.~Avsar, E.~Iancu, L.~McLerran, D.~N.~Triantafyllopoulos,
  JHEP {\bf 0911 } (2009)  105.
  [arXiv:0907.4604 [hep-th]].

\bibitem{Kovchegov:2009du}
  Y.~V.~Kovchegov, S.~Lin,
  JHEP {\bf 1003 } (2010)  057.
  [arXiv:0911.4707 [hep-th]].


\bibitem{Taliotis:2012sx}
  A.~Taliotis,
  arXiv:1212.0528 [hep-th].



\bibitem{Yoshino:2002br}
  H.~Yoshino, Y.~Nambu,
  Phys.\ Rev.\  {\bf D66 } (2002)  065004.
  [gr-qc/0204060].

\bibitem{Kovner:1995ja}
  A.~Kovner, L.~D.~McLerran, H.~Weigert,
  Phys.\ Rev.\  {\bf D52 } (1995)  6231-6237.
  [hep-ph/9502289].



\bibitem{Alver:2008aq}
  B.~Alver, M.~Baker, C.~Loizides, P.~Steinberg,
  [arXiv:0805.4411 [nucl-ex]].

\bibitem{Baier:2006gy}
  R.~Baier and P.~Romatschke,
  Eur.\ Phys.\ J.\ C {\bf 51} (2007) 677
  [nucl-th/0610108].

\bibitem{website}
The full results for this calculation in form of a Mathematica notebook can be found at {\it
https://sites.google.com/site/shockwavecollisionsinads/}


\end{thebibliography}

\end{document}